\documentclass[aps,pra,twocolumn,showpacs,superscriptaddress,10pt,longbibliography]{revtex4-2}

\usepackage[colorlinks = true,linkcolor = red,citecolor = magenta]{hyperref}
\usepackage[sort&compress]{natbib}
\usepackage{scalerel}
\usepackage[normalem]{ulem}
\usepackage{xcolor}
\usepackage{bm}
\usepackage{amsmath}
\usepackage{amsfonts}
\usepackage{amssymb}
\usepackage{graphicx}
\usepackage{subfigure}
\usepackage{mathtools}
\usepackage{physics}
\usepackage{dsfont}
\usepackage{float}
\usepackage{ulem}
\usepackage{lmodern}
\usepackage[many]{tcolorbox}
\usepackage{empheq}
\usepackage{cleveref}
\usepackage{textcomp}
\usepackage{mathrsfs}
\usepackage{bm}
\usepackage[dvipsnames]{xcolor}
\newcommand{\scs}{\scriptscriptstyle}

\newcommand{\smallm}{\scs (\!-\!)}
\newcommand{\smallp}{\scs (\!+\!)}

\newcommand{\smallpm}{\scs (\!\pm\!)}
\begin{document}

\title{Environmentally assisted transport in a diamond chain}
\author{Pavel~S.~Muraev}
\affiliation{Kirensky Institute of Physics, Federal Research Center KSC SB
RAS, 660036, Krasnoyarsk, Russia}
\affiliation{IRC SQC, Siberian Federal University, 660041, Krasnoyarsk, Russia}

\author{Andrey~R.~Kolovsky}
\affiliation{Kirensky Institute of Physics, Federal Research Center KSC SB
RAS, 660036, Krasnoyarsk, Russia}
\affiliation{IRC SQC, Siberian Federal University, 660041, Krasnoyarsk, Russia}

\author{ Dmitrii~N.~Maksimov}
\affiliation{Kirensky Institute of Physics, Federal Research Center KSC SB
RAS, 660036, Krasnoyarsk, Russia}
\affiliation{IRC SQC, Siberian Federal University, 660041, Krasnoyarsk, Russia}

\date{\today}

\begin{abstract}
We study quantum transport in a diamond chain coupled to Markovian reservoirs at its boundaries. The system is subject to uniform dephasing and an external magnetic field threading a flux $\Phi$ per plaquette. We derive exact analytical solutions for $\Phi=0$ and $\Phi=\pi$. We show that in the all-bands-flat regime at $\Phi=\pi$, dephasing destroys Aharonov-Bohm caging, leading to environment-assisted quantum transport. Our results demonstrate that coupling a flat-band system to a dissipative environment provides a viable route for controlling transport in regimes where coherent dynamics alone would result in complete localization.
\end{abstract}

\maketitle

\section{Introduction}

The interplay between lattice geometry and gauge fields provides a powerful route to controlling quantum transport through destructive interference. A paradigmatic realization is the linear diamond (or rhombic) chain threaded by a magnetic flux, which can be implemented as a synthetic gauge field in photonic and atomic systems~\cite{Vidal1998,Longhi2014,Mukherjee2018}. At a gauge-invariant flux of $\Phi=\pi$ per plaquette, tunneling paths connecting neighboring unit cells interfere destructively, yielding an all-bands-flat (ABF) single-particle spectrum and compact localized states (CLS)~\cite{Vidal1998,Lopes2011,Longhi2014,Mukherjee2018,Leykam2018,Lape2025}. This phenomenon, known as Aharonov-Bohm (AB) caging, suppresses ballistic single-particle spreading in the ideal coherent and isolated limit. Recent theoretical work has further generalized ABF constructions with broken time-reversal symmetry, where CLS can support localized circulatory currents controlled by the flux~\cite{Ray2026}.

AB caging thus provides a particularly clear example of localization without disorder. It emerges solely from lattice geometry and quantum interference. Unlike Anderson localization~\cite{Anderson1958}, where disorder suppresses transport through random scattering, AB caging relies on precise phase relations between propagation paths. Consequently, the CLS are tied to coherence and to the caging condition. Perturbations that lift the flat-band degeneracy can unflatten the spectrum and restore transport~\cite{Danieli2024,Danieli2026}. At the same time, the CLS can be robust against certain static perturbations, so the interplay between fine-tuning and protection is nontrivial~\cite{Gligoric2020}.

Before discussing how perturbations affect the caging condition, we briefly recall the broader context of boundary-driven quantum transport. In this setup, two reservoirs at different chemical potentials are connected by a lattice, so a steady current is established~\cite{Landi2022}. For quadratic open systems, i.e. systems with quantum Liouvillians quadratic in field operators, the equation of motion for the covariance matrix closes exactly, and the steady state is Gaussian~\cite{Guo2017,Teretenkov2019,barthel2022}. This makes
boundary-driven tight-binding chains exactly solvable. Furthermore, for a linear chain with dephasing, the exact mean current~\cite{znidaric10,Asadian2013} has been obtained analytically. 

In realistic quantum and photonic systems, coherent dynamics is accompanied by driving and, in interacting platforms, by many-body interactions. These modify the populations and coherences underlying destructive interference, altering the transport blockade of AB caging. Recent experiments on diamond-chain electrical lattices have shown that local driving can selectively excite CLS but also induce instabilities and restore transport when the shifted CLS frequency resonates with a dispersive band~\cite{ChaseMayoral2024}. Theoretically, interactions do not simply restore diffusion, but can also result in many-body flatband localization or quantum caging, restricting transport to bound pairs~\cite{Vidal2000,Danieli2020,Danieli2021}, Interactions in bosonic AB cages can drive a conductance transition between localized and delocalized regimes~\cite{Kolovsky2023}. Superconducting-qubit experiments have observed flat-band localization and interaction-induced delocalization of photons~\cite{Martinez2023}, and Rydberg-atom experiments have revealed the interaction-driven breakdown of AB caging with bound-pair dynamics~\cite{Chen2025}.

Beyond interactions and driving, coupling to an environment introduces dephasing, which plays a dual role. Moderate dephasing can destroy coherent localization and dark states, facilitating transport---environment-assisted quantum transport (ENAQT)~\cite{Mohseni2008,Rebentrost2009}---by driving an anti-Zeno-like escape from localized subspaces~\cite{Kofman2000,Greenfield2026}. ENAQT has been established in disordered networks, excitonic materials, and synthetic quantum hardware~\cite{Dwiputra2021,Alterman2024,Blach2025,Nallamothula2026}. Strong dephasing, by contrast, acts as a continuous measurement, suppressing coherent tunneling and hindering transport---the quantum Zeno effect~\cite{Facchi2008,Greenfield2026}. While ENAQT and the Zeno regime are well studied in disordered networks and photosynthetic complexes, their manifestation in flat-band systems with synthetic flux, and specifically the effect of dephasing on steady-state transport through an AB cage, remains relatively unexplored.

This leads to a fundamental problem of the intricate interplay between coherent interference, external driving, dissipation, and interactions in quantum transport and localization in geometrically frustrated lattices. In this work, we address this question in a boundary-driven diamond chain populated by bosonic particles. The chain is coupled to particle reservoirs at its edges and subject to bulk dephasing. We focus on the non-interacting limit, where the single-particle density matrix
provides an exact description of the steady state. In what follows, we demonstrate how dissipation can either trigger or suppress transport. 

\section{Model}

\begin{figure}
\centering
\includegraphics[width=0.48\textwidth]{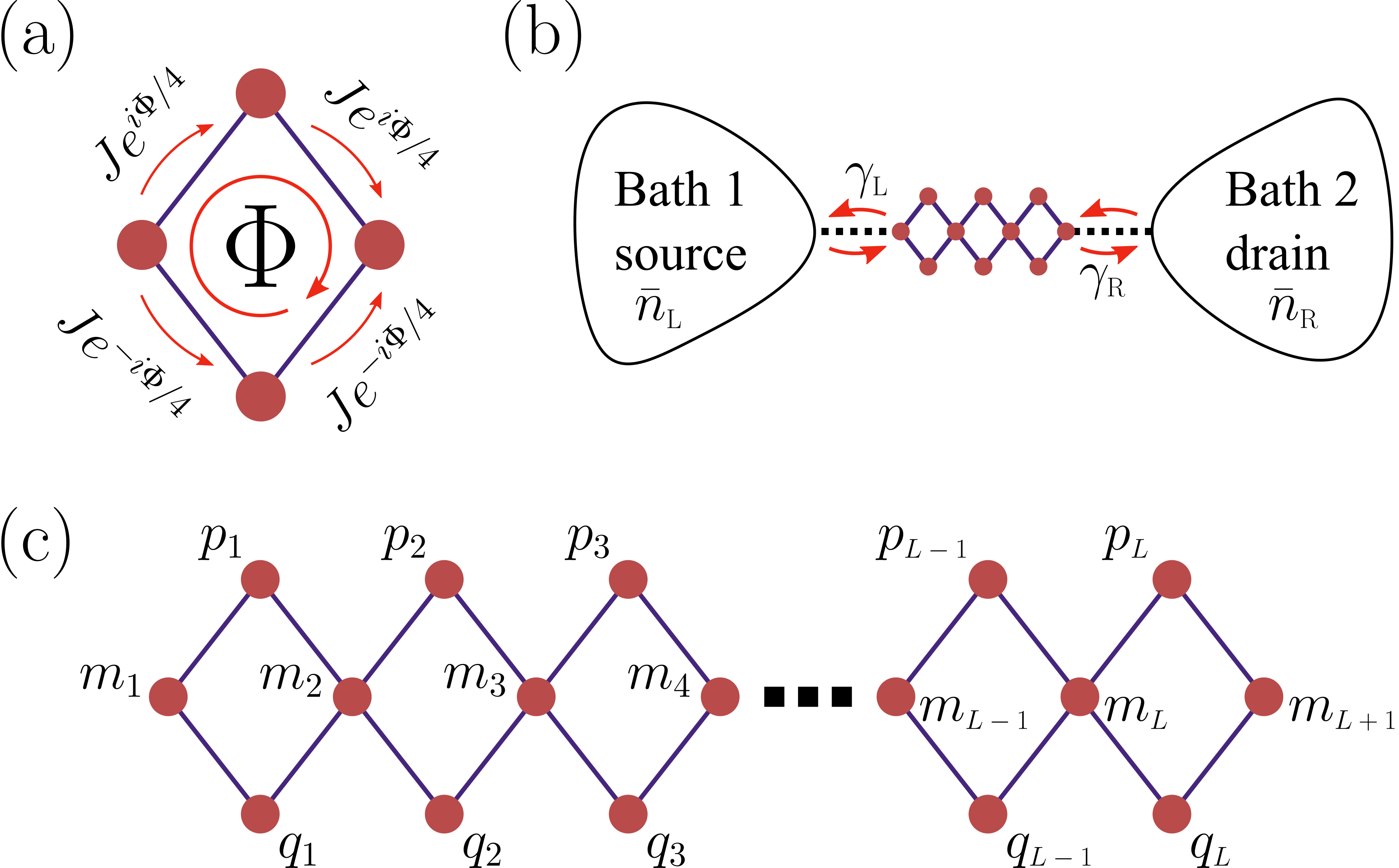}
\caption{A schematic representation of the system. (a) Plaquette of the diamond-chain lattice, emphasizing hopping connections modified by the magnetic flux $\Phi$. (b) Full system configuration with reservoirs. (c) Site-indexing convention.}
\label{fig1}
\end{figure}

We consider a diamond chain of $L$ diamond plaquettes (diamonds) with $M=3L+1$ sites, populated by bosonic particles. The sites are grouped into three sublattices, the lateral sites $p_\ell$, $q_\ell$ with $\ell=1,\dots,L$, and the central sites $m_\ell$, $\ell=1,\dots,L+1$, see Fig.~\ref{fig1}. The tight-binding Hamiltonian reads
\begin{equation}\label{eq:H}
\begin{split}
&\widehat{\mathcal H} = -\frac{J}{2}\sum_{\ell=1}^{L}\Bigl[e^{i\Phi/4}\bigl(\hat a^\dagger_{p_\ell}\hat a_{m_\ell}+\hat a^\dagger_{m_{\ell+1}}\hat a_{p_\ell}\bigr)\\
&+e^{-i\Phi/4}\bigl(\hat a^\dagger_{q_\ell}\hat a_{m_\ell}+\hat a^\dagger_{m_{\ell+1}}\hat a_{q_\ell}\bigr)\Bigr]+\mathrm{h.c.},
\end{split}
\end{equation}
where $\hat{a}^{\dagger}, \hat{a}$ are bosonic creation/annihilation operators, $J$ is the hopping amplitude and $\Phi$ is the magnetic flux per plaquette. The Peierls phases are distributed symmetrically  such that the total flux through the plaquette is $\Phi$. The dynamics is controlled by the Lindblad master equation
\begin{equation}\label{eq:master}
\frac{\partial\widehat{\mathcal R}}{\partial t}
=-i[\widehat{\mathcal H},\widehat{\mathcal R}]
+\!\sum_{\ell=1,{L\!+\!1}}\!\mathcal L_{m_{\ell}}(\widehat{\mathcal R})
+\!\sum_{j\in \{m,p,q\}}\!\mathcal D_j(\widehat{\mathcal R}),
\end{equation}
where $\mathcal L_j$ describes particle exchange with the left ($\ell=1$) and right ($\ell=L+1$) reservoirs,
\begin{equation}\label{eq:res}
\begin{split}
\mathcal L_m(\widehat{\mathcal R})
&=\gamma_m(\bar n_m+1)\Bigl(\hat a_m\widehat{\mathcal R}\hat a_m^\dagger
-\frac12\{\hat a_m^\dagger\hat a_m,\widehat{\mathcal R}\}\Bigr)\\
&+\gamma_m\bar n_m\Bigl(\hat a_m^\dagger\widehat{\mathcal R}\hat a_m
-\frac12\{\hat a_m\hat a_m^\dagger,\widehat{\mathcal R}\}\Bigr),
\end{split}
\end{equation}
with relaxation rate $\gamma_m$ and mean density
of particles in the corresponding reservoir $\bar n_m$ and $\{\ldots,\ldots\}$ stands for the anticommutator.
Throughout, we use $\bar n_{m_1}\equiv\bar n_{\scs{\rm L}}$, $\bar n_{m_{L+1}}\equiv\bar n_{\scs{\rm R}}$, and similarly $\gamma_{m_1}\equiv\gamma_{\scs{\rm L}}$, $\gamma_{m_{L+1}}\equiv\gamma_{\scs{\rm R}}$.
The bulk dephasing is described by
\begin{equation}\label{eq:deph}
\mathcal D_j(\widehat{\mathcal R})
=\kappa\Bigl(\hat n_j\widehat{\mathcal R}\hat n_j
-\frac12\{\hat n_j^2,\widehat{\mathcal R}\}\Bigr),
\end{equation}
where $\kappa$ is the dephasing rate and $\hat n_j=\hat a_j^\dagger\hat a_j$. 

%
\begin{figure*}
\centering
\includegraphics[width=0.8\textwidth]{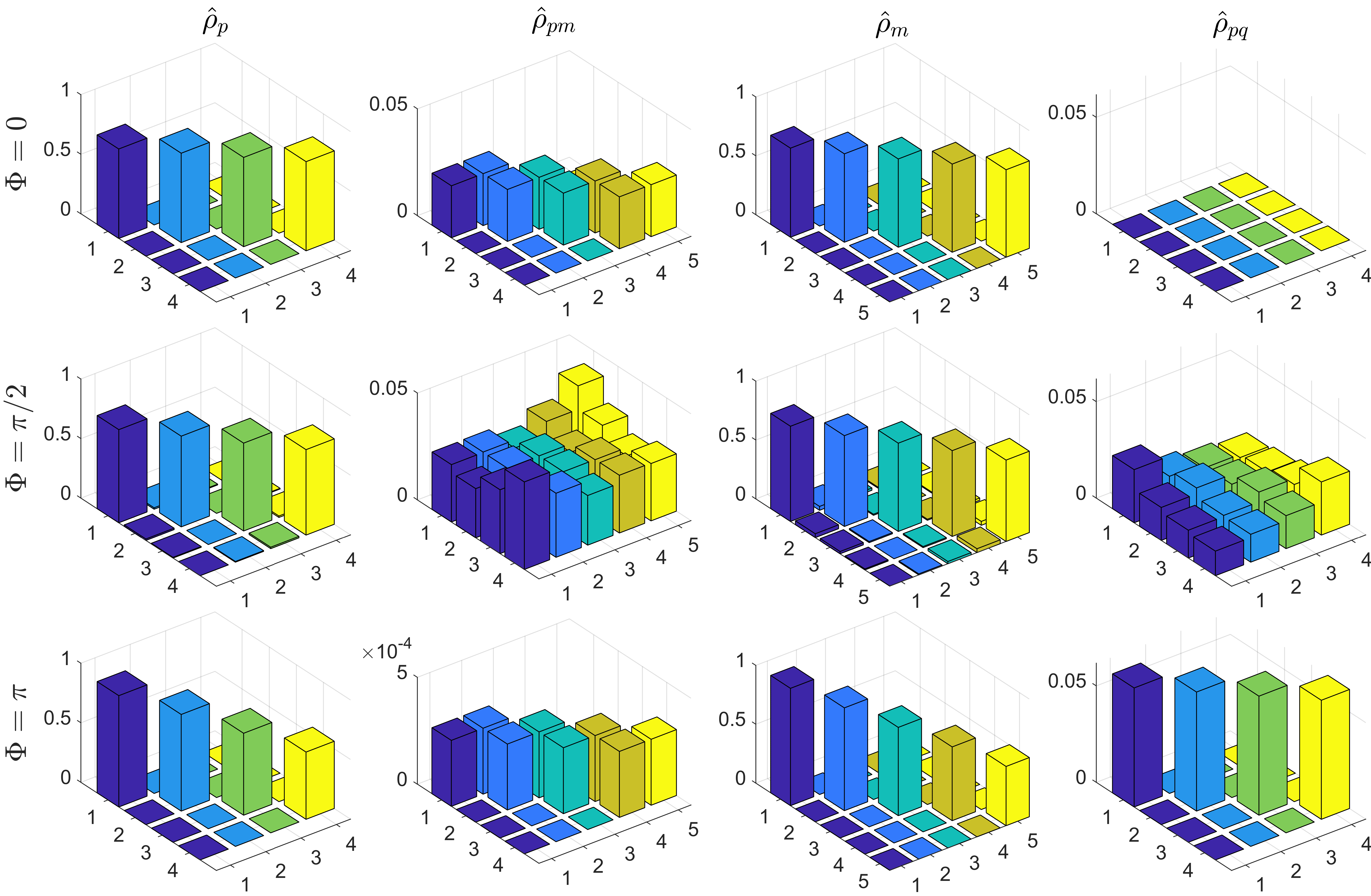}
\caption{Absolute values of the SPDM elements. Columns correspond to the same blocks in Eq.~\eqref{eq:rhoblock}; rows correspond to the same magnetic flux $\Phi$. The  parameters are $J=1$, $L=4$, $\gamma_{\scs{\rm L}}=\gamma_{\scs{\rm R}}=0.2$, $\bar n_{\scs{\rm L}}=1$, $\bar n_{\scs{\rm R}}=0.5$, and $\kappa=0.01$.}
\label{fig2}
\end{figure*}

In the long-time limit, the system reaches a steady state. The two quantities of central interest are the stationary site populations and the stationary particle current. Both follow from the single-particle density matrix (SPDM)
\begin{equation}\label{eq:spdm}
\rho_{\ell', \ell }=\langle\hat a_{\ell}^\dagger\hat a_{\ell'}\rangle
\equiv\mathrm{Tr}\bigl[\hat a_{\ell}^\dagger\hat a_{\ell'}\widehat{\mathcal R}\bigr].
\end{equation}
The equation of motion for the SPDM can be obtained by directly applying the definition to Eq.~\eqref{eq:master}. The result reads
\begin{equation}\label{eq:spdm_eom}
\begin{split}
&\frac{d\hat\rho}{dt}
=-i\bigl(\widehat H_{\mathrm{eff}}\hat\rho-\hat\rho\widehat H_{\mathrm{eff}}^\dagger\bigr)\\
&+\kappa\!\sum_{j\in \{m, p, q\}}\!|j\rangle\langle j|\hat\rho|j\rangle\langle j|
\!+\!\sum_{\ell=1,L\!+\!1}\!\gamma_{m_{\ell}}\bar n_{m_{\ell}}|{m_{\ell}}\rangle\langle {m_{\ell}}|
\end{split}
\end{equation}
with $H_{\mathrm{eff}}$ as the effective non-Hermitian Hamiltonian
\begin{equation}\label{eq:Heff}
\widehat H_{\mathrm{eff}}\!
=\!\widehat H
-i\frac{\kappa}{2}\!\sum_{j\in \{m,p,q\}}\!|j\rangle\langle j|
-\frac{i}{2}\!\sum_{\ell=1,1\!+\!L}\!\gamma_{m_{\ell}}|m_{\ell}\rangle\langle m_{\ell}|.
\end{equation}
The tight-binding single-particle Hamiltonian of the lattice has a block structure in the $p$--$m$--$q$ basis,
\begin{equation}\label{eq:Hblock}
\widehat H=-\frac{J}{2}
\begin{pmatrix}
0 & \widehat D & 0\\
\widehat D^\dagger & 0 & \widehat D^{\mathsf T}\\
0 & \widehat D^* & 0
\end{pmatrix},
\end{equation}
where
\begin{equation}\label{eq:D}
\widehat{D}=e^{-i{\Phi}/{4}}\widehat{J}_{\scs{\smallp}}\!+\!e^{+i{\Phi}/{4}}\widehat{J}_{\scs{\smallm}},
\end{equation}
and $\widehat J_{\scs{\smallpm}}$ are $L\times(L+1)$ matrices with following structure
\begin{align}
\begin{split}
\widehat{J}_{\scs{\smallp}}\!=\!
\left(
\begin{array}{ccccc}
1 & 0 & \cdots & 0 & 0 \\
0 & 1 & \cdots & 0 & 0 \\
\vdots & \vdots & \ddots & \vdots & \vdots  \\
0 & 0 & \cdots & 1 & 0 
\end{array}
\right), \\[8pt]
\widehat{J}_{\scs{\smallm}}\!=\!
\left(
\begin{array}{ccccc}
0 & 1 & 0 & \cdots & 0 \\
0 & 0 & 1 & \cdots & 0 \\
\vdots &\vdots & \vdots & \ddots & \vdots \\
0& 0 & 0 & \cdots & 1 
\end{array}
\right).
\end{split}
\end{align}

Let us write the SPDM in a block form
\begin{equation}
\hat{\rho}=
\left(
\begin{array}{ccc}\label{spdm_block}
\hat{\rho}_{p} & \hat{\rho}_{pm} & \hat{\rho}_{pq} \\
\hat{\rho}_{mp} & \hat{\rho}_{m} & \hat{\rho}_{mq} \\
\hat{\rho}_{qp} & \hat{\rho}_{qm} & \hat{\rho}_{q}
\end{array}
\right),
\end{equation}
and use the following properties of this matrix
\begin{align}\label{prop}
 &\hat{\rho}=\hat{\rho}^{\dagger}, \nonumber \\
&\hat{\rho}=\widehat{P}\hat{\rho}^*\widehat{P},
 \end{align}
where $\widehat{P}$ is given by the equation below 
\begin{equation}\label{eq:P}
\widehat P=
\begin{pmatrix}
0 & 0 & -\widehat{\mathds{1}}_L\\
0 & \widehat{\mathds{1}}_{L\!+\!1} & 0\\
-\widehat{\mathds{1}}_L & 0 & 0
\end{pmatrix}
\end{equation}
with $\widehat{\mathds{1}}_L$ being the $L\!\times\! L$ identity matrix. The second line in Eq.~\eqref{prop} can be proved by performing complex conjugation on the  master equation \eqref{eq:spdm_eom} and using
\begin{equation}
\widehat{H}=-\widehat{P}\widehat{H}^*\widehat{P}.
\end{equation}
Taking into account Eq.~\eqref{prop} one arrives at the SPDM of the form
\begin{equation}\label{eq:rhoblock}
\hat\rho=
\begin{pmatrix}
\hat\rho_p & \hat\rho_{pm} & \hat\rho_{pq}\\
\hat\rho_{pm}^\dagger & \hat\rho_m & -\hat\rho_{pm}^{\mathsf T}\\
\hat\rho_{pq}^* & -\hat\rho_{pm}^* & \hat\rho_p^*
\end{pmatrix},
\end{equation}
where the relations $\hat\rho_m=\hat\rho_m^*$, $\hat\rho_{pq}=\hat\rho_{pq}^{\mathsf T}$, and $\hat\rho_p=\hat\rho_p^*$ follow from Hermiticity and symmetry $\hat\rho=\widehat P\hat\rho^*\widehat P$.


The particle current through the chain can be calculated as
\begin{equation}\label{eq:current}
I=\gamma_{\scs{\rm L}}(\bar n_{\scs{\rm L}}-n_{m_1})
=\gamma_{\scs{\rm R}}(n_{m_{L+1}}-\bar n_{\scs{\rm R}}),
\end{equation}
where $n_{m_1}\equiv\rho_{m_1m_1}$ and $n_{m_{L+1}}\equiv\rho_{m_{L+1}m_{L+1}}$ are the populations at the edge-sites. Notice that the equality for two definitions of the currents in Eq.~\eqref{eq:current} holds only for the stationary case, $d\hat\rho/dt=0$, in which Eq.~\eqref{eq:spdm_eom} reduces to a closed linear system for the elements of four unique blocks of the SPDM. An example numerical solution for the SPDM blocks is demonstrated in Fig.~\ref{fig2} for $\Phi = 0,\:\pi/2,\:\pi$ at steady state. Note that at $\Phi=0,\pi$ all blocks have a diagonal form. All square matrices are diagonal, whereas the rectangular $\hat{\rho}_{pm}$ has two non-zero main sub-diagonals. This can be contrasted to $\Phi={\pi}/{2}$ where $\hat{\rho}_{pm}$ and $\rho_{pg}$ are dense matrices. Our numerical simulations reveal that $\Phi=0, \pi$ are the only specific cases when the off-diagonal long range order is suppressed without dephasing. In general $\Phi\in(0, \pi)$ the SPDM blocks are dense matrices.

\nopagebreak

\section{Solutions for current}\label{Sec:IV}
\subsection{$\pi$-flux: Aharonov-Bohm caging and dephasing-assisted transport}

At $\Phi=\pi$, the steady-state equations are solved by
\begin{align}\label{eq:Ansatz_pi}
    \begin{split}
        \hat\rho_{pm} &=a\widehat D^*,\\
        \hat\rho_{pq} &=ib\,\widehat{\mathds{1}}_L,\\
        \hat\rho_m    &=\mathrm{diag}(n_{m_1},\dots,n_{m_{L+1}}),\\
        \hat\rho_p    &=\mathrm{diag}(n_{p_1},\dots,n_{p_L}),
    \end{split}
\end{align}
where $a$ and $b$ are unknown real numbers. After substituting the above Ansatz into the master equation~\eqref{eq:spdm_eom} with $\partial{\hat{\rho}}/\partial t=0$, one arrives at the following expressions
\begin{align}\label{Eqset_pi}
& n_{p_{\ell}}-n_{m_{\ell}}+2\frac{\kappa^2+J^2}{\kappa J}a+\gamma_{\scs{\rm L}}\frac{a}{J}\delta_{\ell,1}=0, \nonumber \\
& n_{m_{\ell+1}}-n_{p_{\ell}}+2\frac{\kappa^2+J^2}{\kappa J}a+\gamma_{\scs{\rm R}}\frac{a}{J}\delta_{\ell,L}=0, \nonumber \\
& \gamma_{\scs{\rm L}}n_{m_1}+2{J}a=\gamma_{\scs{\rm L}}\bar{n}_{\scs{\rm L}}, \nonumber \\
& \gamma_{\scs{\rm R}}n_{m_L}-2{J}a=\gamma_{\scs{\rm R}}\bar{n}_{\scs{\rm R}},
 \end{align}
at the same time the parameter $b$ is found as
\begin{equation}
b = -\frac{2J}{\kappa}a. 
\end{equation}
Details on deriving the above equation are provided in Appendix~\ref{App:A}.
Equation~\eqref{Eqset_pi} can be solved for $a$. The result reads
\begin{equation}\label{eq:a_pi}
    a = \frac{\kappa J \gamma_{\scs{\rm L}}\gamma_{\scs{\rm R}}(\bar{n}_{\scs{\rm L}}-\bar{n}_{\scs{\rm R}})}{\kappa (2J^2+\gamma_{\scs{\rm L}}\gamma_{\scs{\rm R}})(\gamma_{\scs{\rm L}}+\gamma_{\scs{\rm R}})+4\gamma_{\scs{\rm L}}\gamma_{\scs{\rm R}} L(\kappa^2+J^2)} .
\end{equation}
Populations of the edge sites $n_{m_1}$ and $n_{m_{L+1}}$ can be found from last two lines of Eq.~\eqref{Eqset_pi} and all the other unknown quantities can be found recurrently with application of the first two lines in Eq.~\eqref{Eqset_pi}.
The expression for the probability current can be found from~\eqref{eq:current}, so the final result reads
\begin{equation}\label{eq:I_pi}
I_\pi=\frac{2\kappa J^2\gamma_{\scs{\rm L}}\gamma_{\scs{\rm R}}(\bar n_{\scs{\rm L}}-\bar n_{\scs{\rm R}})}
{\kappa(2J^2+\gamma_{\scs{\rm L}}\gamma_{\scs{\rm R}})(\gamma_{\scs{\rm L}}+\gamma_{\scs{\rm R}})
+4\gamma_{\scs{\rm L}}\gamma_{\scs{\rm R}} L(\kappa^2+J^2)}.
\end{equation}
Let us examine Eq.~\eqref{eq:I_pi}. In the coherent limit $\kappa\to0$, the current vanishes linearly in $\kappa$,
\begin{equation}\label{eq:I_pi_small}
I_\pi\simeq \frac{\kappa(\bar n_{\scs{\rm L}}-\bar n_{\scs{\rm R}})}{2L}\to0.
\end{equation}
This is the AB caging regime. Destructive interference suppresses transport despite the finite reservoir bias. As $\kappa$ increases, the current rises, reaches a maximum, and then decreases as
\begin{equation}\label{eq:I_pi_large}
I_\pi\simeq \frac{J^2(\bar n_{\scs{\rm L}}-\bar n_{\scs{\rm R}})}{2L\kappa},
\qquad \kappa\gg J.
\end{equation}
The maximum occurs at
\begin{equation}\label{eq:kappa_opt}
\kappa=J,
\end{equation}
with
\begin{equation}\label{eq:I_max}
I_\pi^{\max}
=\frac{2J^2\gamma_{\scs{\rm L}}\gamma_{\scs{\rm R}}(\bar n_{\scs{\rm L}}-\bar n_{\scs{\rm R}})}
{(2J^2+\gamma_{\scs{\rm L}}\gamma_{\scs{\rm R}})(\gamma_{\scs{\rm L}}+\gamma_{\scs{\rm R}})+8\gamma_{\scs{\rm L}}\gamma_{\scs{\rm R}} LJ}.
\end{equation}
For $L\gg1$ one obtains 
\begin{equation}
I_\pi^{\max}\simeq J\frac{\bar n_{\scs{\rm L}}-\bar n_{\scs{\rm R}}}{4L}.
\end{equation}
The nonmonotonic dependence of $I_\pi$ on $\kappa$ is the hallmark of ENAQT. Moderate dephasing breaks the destructive interference responsible for AB caging and restores transport. At strong dephasing, the current is suppressed again, consistent with the quantum Zeno mechanism. 

\subsection{Zero magnetic flux: ballistic-to-diffusive crossover}

At $\Phi=0$, $\widehat D=\widehat J_{\smallp}+\widehat J_{\smallm}$. The steady-state equations are solved by the Ansatz
\begin{align}\label{eq:Ansatz_0}
    \begin{split}
        \hat\rho_{pm} &=ia(\widehat J_{\smallp}-\widehat J_{\smallm}), \\
        \hat\rho_{pq} &=0, \\
        \hat\rho_m    &=\mathrm{diag}(n_{m_1},\dots,n_{m_{L+1}}),\\
        \hat\rho_p    &=\mathrm{diag}(n_{p_1},\dots,n_{p_L}).
    \end{split}
\end{align}
After the same steps as in previous subsection one arrives at
\begin{equation}\label{eq:a0}
a=\frac{J\gamma_{\scs{\rm L}}\gamma_{\scs{\rm R}}(\bar n_{\scs{\rm L}}-\bar n_{\scs{\rm R}})}
{(2J^2+\gamma_{\scs{\rm L}}\gamma_{\scs{\rm R}})(\gamma_{\scs{\rm L}}+\gamma_{\scs{\rm R}})+4\gamma_{\scs{\rm L}}\gamma_{\scs{\rm R}} L\kappa}.
\end{equation}
The corresponding current is
\begin{equation}\label{eq:I0}
I_0=\frac{2J^2\gamma_{\scs{\rm L}}\gamma_{\scs{\rm R}}(\bar n_{\scs{\rm L}}-\bar n_{\scs{\rm R}})}
{(2J^2+\gamma_{\scs{\rm L}}\gamma_{\scs{\rm R}})(\gamma_{\scs{\rm L}}+\gamma_{\scs{\rm R}})+4\gamma_{\scs{\rm L}}\gamma_{\scs{\rm R}} L\kappa}.
\end{equation}
Equation~\eqref{eq:I0} is maximal at $\kappa=0$ and decreases monotonically with increasing dephasing. This is the expected behavior for a system without caging. Coherent transport is already efficient, and dephasing only disrupts it. 
It is worth mentioning that Equation~\eqref{eq:I0} is analogous to the solutions for the ballistic-to-diffusive crossover found for boundary-driven bosonic chains with bulk dephasing~\cite{Asadian2013}. At $\kappa=0$ the current is independent of the length of the chain, but for $\kappa\neq 0$ the current is asymptotically inversely dependent on the chain length.

\subsection{Arbitrary magnetic flux}

For arbitrary $\Phi$, the steady-state equations generally do not admit a simple closed-form diagonal Ansatz due to the off-diagonal long range order. In the strong-dephasing limit $\kappa\gg J$, the off-diagonal elements must nonetheless be damped, and following substitutions can be assumed for SPDM blocks
\begin{align}\label{sub1}
\hat{\rho}_{pm}=
\left(
\begin{array}{ccccc}
a & b & \cdots & \cdots & \cdots  \\
\cdots & a & b & \cdots & \cdots \\
\vdots & \vdots & \vdots & \vdots & \vdots \\
\cdots & \cdots & \cdots & a & b
\end{array}
\right),
\end{align}
and
\begin{align}\label{sub2}
\hat{\rho}_{pq}=
\left(
\begin{array}{cccc}
c & \cdots & \cdots & \cdots  \\
\cdots & c & \cdots & \cdots \\
\vdots & \vdots & \vdots & \vdots \\
\cdots & \cdots & \cdots & c
\end{array}
\right).
\end{align}
For blocks $\hat{\rho}_m$ and $\hat\rho_{p}$ substitutions are in the same form as in~\eqref{eq:Ansatz_pi} and~\eqref{eq:Ansatz_0}. Using substitutions above one arrives at the following result for stationary current
\begin{equation}\label{eq:I_gen}
I\!=\!\frac{2\kappa J^2\gamma_{\scs{\rm L}}\gamma_{\scs{\rm R}}(\bar n_{\scs{\rm L}}-\bar n_{\scs{\rm R}})}
{\kappa(2J^2\!+\!\gamma_{\scs{\rm L}}\gamma_{\scs{\rm R}})(\gamma_{\scs{\rm L}}\!+\!\gamma_{\scs{\rm R}})
\!+\!4\gamma_{\scs{\rm L}}\gamma_{\scs{\rm R}} L(\kappa^2\!+\!FJ^2)},
\end{equation}
with
\begin{align}\label{eq:F}
    \begin{split}
        F &=\sin^2\!\Bigl(\frac{\Phi}{2}\Bigr)
        +\frac{\bar a}{2}\sin\Phi, \\
        \bar a &=-\frac{2J^2L\sin\Phi}
        {\kappa(\gamma_{\scs{\rm L}}+\gamma_{\scs{\rm R}})+4\kappa^2L
        +4J^2L\cos^2(\Phi/2)}.
    \end{split}
\end{align}
Details of the derivation of Eq.~\eqref{eq:I_gen} are provided in Appendix~\ref{App:B}. This expression is exact at $\Phi=0$ and $\Phi=\pi$ for all $\kappa$, where it reduces to Eqs.~\eqref{eq:I_pi}
and~\eqref{eq:I0}, respectively. For intermediate fluxes it provides a leading-order approximation valid for $\kappa\gtrsim J$. In the opposite limit, $\kappa\ll J$, the proposed approximation
does not capture the full structure of the steady-state density matrix. Numerically, the current at intermediate fluxes is well described by $I\simeq I_0\cos^2(\Phi/2)$, consistent with
the empirical relation in ~\cite{muraev2021} for non-interacting bosons in the same lattice without dephasing. 

\subsection{Numerical examples}



\begin{figure}[th]
\centering
\includegraphics[width=.9\linewidth]{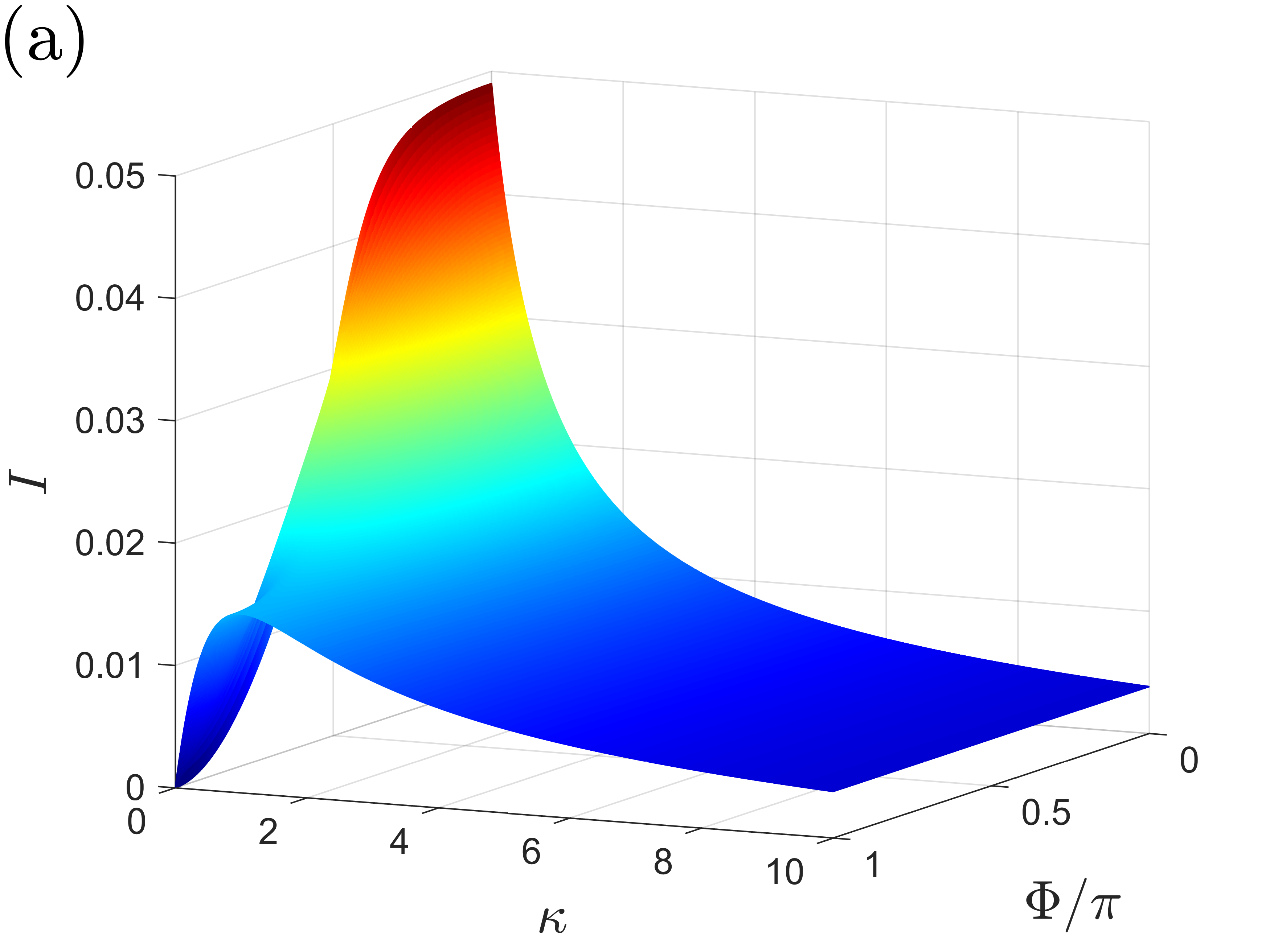}
\includegraphics[width=.9\linewidth]{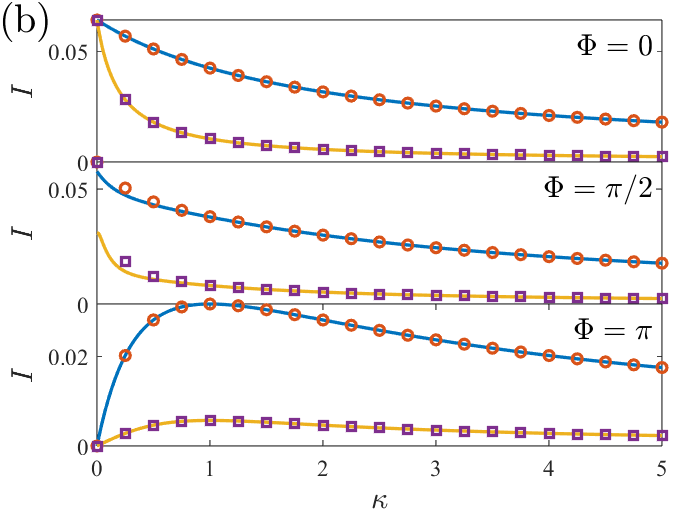}
\caption{(a) Probability current as a function of magnetic flux $\Phi$ and dephasing rate $\kappa$. The parameters are $L=6$, $J=1$, $\gamma_{\scs{\rm L}}=\gamma_{\scs{\rm R}}=0.2$, $\bar n_{\scs{\rm L}}=1$, and $\bar n_{\scs{\rm R}}=0.5$. (b) Probability current as a function of dephasing $\kappa$. Numerical results (solid lines) and analytical results (markers) are shown. Blue line and circles correspond to $L=2$; yellow line and squares correspond to $L=20$. Results for $\Phi=0,\;\pi/2,\;\pi$. The parameters are $J=1$, $\gamma_{\scs{\rm L}}=0.2$, $\gamma_{\scs{\rm R}}=0.4$, $\bar n_{\scs{\rm L}}=1$, and $\bar n_{\scs{\rm R}}=0.5$.}
\label{fig3_4}
\end{figure}

Figure~\ref{fig3_4}~(a) shows numerical results for the steady-state current as a function of $\Phi$ and $\kappa$ for $L=6$. The ENAQT region is visible around $\Phi=\pi$ and $\kappa\simeq J$, while at $\Phi=0$ the current decreases monotonically with $\kappa$. Figure~\ref{fig3_4}~(b) compares the numerical solution of the Eq.~\eqref{eq:spdm_eom} in steady state with the analytic results of Eqs.~\eqref{eq:I_pi} and~\eqref{eq:I0} for $L=2$ and $L=20$. The agreement is ideal at $\Phi=0$ and $\Phi=\pi$ for all $\kappa$. At $\Phi=\pi/2$, the analytic solution deviates from the numerical curves for $\kappa \lesssim J$, as expected from the approximate nature of the strong-dephasing Ansatz at intermediate fluxes.

\section{Conclusion}
We have studied steady-state transport through a boundary-driven diamond chain subject to bulk dephasing and an external magnetic field. The exact analytic solutions for the probability current are derived for flux values $\Phi=0,\pi$.  It is shown that the Aharonov-Bohm caging at  magnetic flux values $\Phi=\pi$ per plaquette is lifted by moderate dephasing,
which restores transport with the optimal dephasing rate $\kappa=J$ that is independent of the reservoir coupling rates and of the number of sites in the chain. It is demonstrated that at $\Phi=0$ dephasing only suppresses the current. These results establish the diamond chain as a minimal lattice realization of
environment-assisted quantum transport and as a flux-controlled switch between dephasing-suppressed and dephasing-assisted regimes.

The physical picture that emerges is the following. In the coherent limit, the chain spectrum at $\Phi=\pi$ is dispersionless, and transport through the chain is blocked by destructive interference.
Coupling the chain to reservoirs at the edges does not by itself result in transport, because the caging condition is protected by the lattice geometry. Dephasing breaks this protection. At moderate
dephasing rates, the noise destroys the phase relations responsible for caging and enables transport. The optimal dephasing rate $\kappa=J$ is reminiscent of the ``Goldilocks'' condition discussed in the context of excitonic transport~\cite{Alterman2024,Blach2025}, where the optimal noise strength is comparable to the relevant coherent energy scale. 
In contrast, at $\Phi=0$ the chain is a tight-binding lattice with a dispersive band. Ballistic transport is efficient, and dephasing only degrades it, consistent with the ballistic-to-diffusive crossover known for open chains with dephasing~\cite{Asadian2013,znidaric10}. 

Several directions for future work follow naturally. First, the interplay between dephasing and interactions could be explored along the lines of ~\cite{Danieli2020,Danieli2021,Kolovsky2023,Chen2025}. Second, the role of non-Markovian noise and structured reservoirs could be investigated ~\cite{Rebentrost2009,Dwiputra2021}. Third, the predicted environmentally assisted transport peak at $\kappa=J$ could be tested in photonic waveguide arrays~\cite{Mukherjee2018,Longhi2014}, cold-atom synthetic lattices,
or electrical-circuit flat-band lattices~\cite{ChaseMayoral2024,Lape2025}.

{\em Acknowledgments.} 
This work was supported by the Russian Science Foundation under Grant No. 25-12-00268.

\bibliography{mybib}
\appendix
\section{Steady-state equations for the SPDM blocks}\label{App:A}

Substituting the block form of the SPDM given by Eq.~\eqref{eq:rhoblock} into the steady-state master equation~\eqref{eq:spdm_eom} results in the system of four matrix equations
\begin{widetext}
\begin{align}
&i\frac{J}{2}\bigl(\widehat D\hat\rho_{pm}^\dagger
-\hat\rho_{pm}\widehat D^\dagger\bigr)
-\kappa\bigl[\hat\rho_p-\mathrm{diag}(\hat\rho_p)\bigr]=0,
\label{eq:A1}\\
&i\frac{J}{2}\bigl(\widehat D^\dagger\hat\rho_p
+\widehat D^{\mathsf T}\hat\rho_{pq}^*
-\hat\rho_m\widehat D^\dagger\bigr)
-\kappa\hat\rho_{pm}^\dagger
-\frac12\bigl(\gamma_{\scs{\rm R}}|m_{L+1}\rangle\langle m_{L+1}|\hat\rho_{pm}^\dagger
+\gamma_{\scs{\rm L}}|m_1\rangle\langle m_1|\hat\rho_{pm}^\dagger\bigr)=0,
\label{eq:A2}\\
&i\frac{J}{2}\bigl(\widehat D^*\hat\rho_{pm}^\dagger
+\hat\rho_{pm}^*\widehat D^\dagger\bigr)
-\kappa\hat\rho_{pq}^*=0,
\label{eq:A3}\\
&i\frac{J}{2}\bigl(\widehat D^\dagger\hat\rho_{pm}
-\widehat D^{\mathsf T}\hat\rho_{pm}^*
+\hat\rho_{pm}^{\mathsf T}\widehat D^*
-\hat\rho_{pm}^\dagger\widehat D\bigr)
-\frac{\gamma_{\scs{\rm L}}}{2}\bigl(|m_1\rangle\langle m_1|\hat\rho_m
-2\bar n_{\scs{\rm L}}|m_1\rangle\langle m_1|
+\hat\rho_m|m_1\rangle\langle m_1|\bigr)
\nonumber\\
&\qquad-\frac{\gamma_{\scs{\rm R}}}{2}\bigl(|m_{L+1}\rangle\langle m_{L+1}|\hat\rho_m
-2\bar n_{\scs{\rm R}}|m_{L+1}\rangle\langle m_{L+1}|
 +\hat\rho_m|m_{L+1}\rangle\langle m_{L+1}|\bigr)-\kappa\bigl[\hat\rho_m-\mathrm{diag}(\hat\rho_m)\bigr]=0.
\label{eq:A4}
\end{align}
\end{widetext}
The Ans\"atze~\eqref{eq:Ansatz_pi} and~\eqref{eq:Ansatz_0} solve these equations in the two limiting flux cases. For the $\pi$-flux case, the identities
 \begin{align}\label{eq:J_ind}
 & \widehat{J}_{\scs{\smallpm}}\widehat{J}_{\scs{\smallpm}}^{\dagger}=\widehat{\mathds{1}}_L, \nonumber \\
 &
 \widehat{J}_{\scs{\smallp}}^{\dagger}\widehat{J}_{\scs{\smallp}}=\widehat{\mathds{1}}_{L\!+\!1}- |{L\!+\!1}\rangle\langle {L\!+\!1}|, \nonumber \\
 &
 \widehat{J}_{\scs{\smallm}}^{\dagger}\widehat{J}_{\scs{\smallm}}=\widehat{\mathds{1}}_{L\!+\!1}- |{1}\rangle\langle {1}|.
 \end{align}
are used to reduce Eqs.~\eqref{eq:A1}--\eqref{eq:A4} to the closed system for $a$, $b$, and the site populations quoted in the main text.

\section{Current derivation for arbitrary magnetic flux}\label{App:B}

Equations~\eqref{eq:A1}~and~\eqref{eq:A4} represent square matrices of  size $L\!\times\!L$ and $L+\!1\!\times\!L+\!1$, respectively. One can see through a direct substitution of the Eq.~\eqref{sub1} and Eq.~\eqref{sub2} that the main diagonals of the resulting two equations turn into identity if we
chose
\begin{align}\label{first_two}
a=e^{-i\Phi/4}\frac{\bar{a}+i}{2J}{I}, \nonumber \\
b=e^{+i\Phi/4}\frac{\bar{b}-i}{2J}{I},
\end{align}
where $\bar{a}$, $\bar{b}$ are real numbers, and $I$ is the probability current across the lattice. The probability current is introduced into Eq.~\eqref{first_two} to be consistent with the~\eqref{eq:current}. Equation~\eqref{eq:A3} is also a square matrix. Using Eq.~\eqref{sub1} and Eq.~\eqref{sub2} for the main diagonal of Eq.~\eqref{eq:A3} one finds
\begin{equation}\label{c}
c=-i\frac{I}{2\kappa}\left[e^{-i\Phi/2}\bar{a}+e^{+i\Phi/2}\bar{b}+2\sin\left(\frac{\Phi}{2}\right)\right]
\end{equation}
Equation~\eqref{eq:A2} is a $L+\!1\!\times\! L$ rectangular matrix. The first two sub-diagonals yield
\begin{widetext}

\begin{align}
        & n_{p_1} = n_{m_1}+\langle m_1 |\hat{\rho}_{m}| m_2 \rangle e^{-i\Phi/2}-c^*e^{-i\Phi/2}-i\frac{a^*}{J}(2\kappa+\gamma_1)e^{-i\Phi/4}, \nonumber \\
        & n_{m_{\ell+1}} = n_{p_{\ell}}-\langle m_{\ell+\!1} |\hat{\rho}_{m}| m_{\ell} \rangle e^{+i\Phi/2}+\langle p_{\ell+\!1} |\hat{\rho}_{p}| p_{\ell} \rangle e^{+i\Phi/2}+c^*e^{+i\Phi/2}+\langle p_{\ell+\!1} |\hat{\rho}_{pq}^*| q_{\ell} \rangle+i\frac{2b^*}{J}\kappa e^{+i\Phi/4}, \ \ell=1,2,\ldots, L\!-\!1, \nonumber \\
        & n_{p_{\ell}} = n_{m_{\ell}}+\langle m_{\ell} |\hat{\rho}_{m}| m_{\ell+\!1} \rangle e^{-i\Phi/2}-\langle p_{\ell-\!1} |\hat{\rho}_{p}| p_{\ell} \rangle e^{-i\Phi/2}-c^*e^{-i\Phi/2}-\langle p_{\ell-\!1} |\hat{\rho}_{pq}^*| q_{\ell} \rangle-i\frac{2a^*}{J}\kappa e^{-i\Phi/4}, \ \ell=2,3,\ldots, L, \nonumber \\
        & n_{m_{L+1}} = n_{p_{L}}-\langle m_{L+\!1} |\hat{\rho}_{m}| m_L \rangle e^{+i\Phi/2}+c^*e^{+i\Phi/2}+i\frac{b^*}{J}(2\kappa+\gamma_2)e^{+i\Phi/4}.
\end{align}
\end{widetext}
The above equations can be simplified with the following substitutions 
\begin{align}
    \begin{split}
        \hat\rho_m    &=\mathrm{diag}(n_{m_1},\dots,n_{m_{L+1}}),\\
        \hat\rho_p    &=\mathrm{diag}(n_{p_1},\dots,n_{p_L}).
    \end{split}
\end{align}
These substitutions lead to two results. First, they give the condition $\overline{a}=\overline{b}$ in Eq.~\eqref{first_two}, which is a consequence of the diagonal structure of $\hat\rho_p$ and
$\hat\rho_m$ rather than an independent approximation. Second, they produce the recurrent solution linking the populations at the edge sites
\begin{align}
\begin{split}
 n_{m_{L+1}} &=n_{m_1}+{\rm Re}\Bigl[L\left(c-c^*\right)e^{-i\Phi/2}   \\[8pt]
 &-i\frac{\gamma_1 a^*+\gamma_2 b}{J}e^{-i\Phi/4} 
 -2i\kappa L\frac{a^*+ b}{J}e^{-i\Phi/4}\Bigl] 
\end{split}
\end{align}
The above equation together with Eq.~\eqref{first_two}, Eq.~\eqref{eq:current}, and Eq.~\eqref{c} result in the formula~\eqref{eq:I_gen} for the probability current. One can expand expression~\eqref{eq:I_gen} in a next series in powers of $1/\kappa$
\begin{align}
\begin{split}
  &I/C_0 =   \frac{1}{\kappa} - B\frac{1}{\kappa ^2} + \left[B^2 - J^2\sin^2{\left(\frac{\Phi}{2}\right)}\right]\frac{1}{\kappa ^3} - \\
  &\left[B^3 -2J^2B\sin^2{\left(\frac{\Phi}{2}\right)}\right]\frac{1}{\kappa^4} + \\
  &\left[B^4 - 3B^2J^2\sin^2{\left(\frac{\Phi}{2}\right)} + \frac{J^4\sin^2{\Phi}}{4} + J^4\sin^4{\left(\frac{\Phi}{2}\right)}\right]\frac{1}{\kappa^5} ,\\[8pt]
  &C_0 =\frac{J^2 (\bar{n}_1 - \bar{n}_L)}{2 L}, \\
  &B = \frac{(2J^2 + \gamma_1 \gamma_L)(\gamma_1 + \gamma_L)}{4 \gamma_1 \gamma_L L}.
\end{split}
\end{align}

\end{document}